\documentclass[journal]{IEEEtran}
\usepackage{amsmath,amsfonts}
\usepackage{algorithmic}
\usepackage{algorithm}
\usepackage{array}
\usepackage[caption=false,font=normalsize,labelfont=sf,textfont=sf]{subfig}
\usepackage{textcomp}
\usepackage{stfloats}
\usepackage{url}
\usepackage{verbatim}
\usepackage{graphicx}
\usepackage{cite}
\usepackage{amsmath}
\usepackage{xcolor}

\begin{document}

\title{Power efficient ReLU design for neuromorphic computing using spin Hall effect}

\author{\IEEEauthorblockN{Venkatesh Vadde$^{1,}$\IEEEauthorrefmark{1},
Bhaskaran Muralidharan$^{1,}$\IEEEauthorrefmark{2},
Abhishek Sharma$^{2,}$\IEEEauthorrefmark{3}}

\IEEEauthorblockA{$^{1}$Department of Electrical Engineering, Indian Institute of Technology Bombay, Powai, Mumbai-400076, India}
\IEEEauthorblockA{$^{2}$Department of Electrical Engineering, Indian Institute of Technology Ropar, Rupnagar, Punjab-140001, India}

\thanks{\IEEEauthorrefmark{1}vaddevenkatesh19@gmail.com,
\IEEEauthorrefmark{2}bm@ee.iitb.ac.in,
\IEEEauthorrefmark{3}abhishek@iitrpr.ac.in}}
\maketitle

\begin{abstract}
We demonstrate a magnetic tunnel junction injected with spin Hall current to exhibit linear rotation of magnetization of the free-ferromagnet using only the spin current. Using the linear resistance change of the MTJ, we devise a circuit for the rectified linear activation (ReLU) function of the artificial neuron. We explore the role of different spin Hall effect (SHE) heavy metal layers on the power consumption of the ReLU circuit. We benchmark the power consumption of the ReLU circuit with different SHE layers by defining a new parameter called the spin Hall power factor. It combines the spin Hall angle, resistivity, and thickness of the heavy metal layer, which translates to the power consumption of the different SHE layers during spin-orbit switching/rotation of the free FM. We employ a hybrid spintronics-CMOS simulation framework that couples Keldysh non-equilibrium Green's function formalism with Landau-Lifshitz-Gilbert-Slonzewski equations and the HSPICE circuit simulator to account for diverse physics of spin-transport and the CMOS elements in our proposed ReLU design.
We also demonstrate the robustness of the proposed ReLU circuit against thermal noise and non-trivial power-error trade-off that enables the use of an unstable free-ferromagnet for energy-efficient design.
Using the proposed circuit, we evaluate the performance of the convolutional neural network for MNIST datasets and demonstrate comparable classification accuracies to the ideal ReLU with an energy consumption of 75 $pJ$ per sample.

\end{abstract}


\section{Introduction}
Artificial neural networks (ANNs) are widely used by machine learning and data science communities to solve complex problems. The ANNs are inspired by the biological brains, which have memory and computing intertwined to solve diverse problems while consuming low energy \cite{markovic2020physics, schuman2017survey}. The von Neumann-based modern computers that separate memory and computing are not suitable for hardware implementation of neural networks \cite{grollier2020neuromorphic}. Neural networks contain highly interconnected perceptrons, which define the mathematical model of biological neurons as the sum of weighted inputs passed through an activation function \cite{taud2018multilayer}.

Learning in neural networks can be achieved by using activation functions \cite{ide2017improvement}. The activation function introduces non-linearity to the network and enables the network to learn complex data structures and differentiate between outputs. Traditionally, sigmoid and tanh activation functions have been widely utilized. But these standard functions limit the network's ability to learn since they saturate when the input is very high or very low\cite{goodfellow2016deep,glorot2011deep}. The sigmoid and tanh functions also face the vanishing gradient problem, where the gradient information used to learn networks becomes almost zero for deep networks, thus affecting the deep network's learning capacity\cite{goodfellow2016deep}.
Glorot et al.\cite{glorot2011deep} showed that the rectified linear unit (ReLU) activation function can improve the learning speed of various neural networks. The ReLU function also overcomes the vanishing gradient and saturation problems that tanh and sigmoid functions face. It often produces better results than traditional functions such as tanh and sigmoid in neural networks \cite{glorot2011deep,goodfellow2016deep}. Thus, the ReLU has become a default activation function for various neural networks \cite{nair2010rectified,jarrett2009best,ide2017improvement}.

The ReLU function is described as
\[
f(x) =
\begin{cases}
0 & \text{if $x\le 0$} \\
x & \text{if $x> 0$}
\end{cases}
\]

The activation function using CMOS technology has been explored by a few works \cite{chang2019hardware, geng2020analog, priyanka2018cmos}, but these realizations of the activation function require additional interconnect circuits to interface with synaptic and max-pooling layers\cite{vadde2022orthogonal} of the neural network. The CMOS implementation is also limited by the area and energy requirements \cite{grollier2020neuromorphic, indiveri2011neuromorphic}. On the other hand, spintronics provides a wide range of devices that can be engineered to have non-volatility, plasticity, oscillatory and stochastic behavior\cite{datta2018lessons, apalkov2016magnetoresistive, sharma2017resonant, camsari2017implementing, hirohata2020review,sengupta2018neuromorphic}. These properties suit well for in-memory computing, enabling neuromorphic computing and taking advantage of the paradigm ``let physics do the computing''\cite{parihar2017computing}. In this paper, we demonstrate an MTJ-based design to emulate the ReLU function, which can be easily connected to cross-bar-based synaptic layers\cite{jo2010nanoscale,fan2014design} and the max-pooling layers\cite{vadde2022orthogonal}.

Current induced spin-orbit torques (SOTs) originating from spin Hall effect\cite{song2021spin,liu2012current,takahashi2008spin,hirsch1999spin} in heavy metal(HM)/ferromagnet(FM) hetero-structure, have recently emerged as an energy-efficient manipulation of magnetization at the nanoscale.
The efficient conversion of the charge current to the spin current via SHE is quantified by the spin Hall angle($\theta_{SH}$). There have been consistent efforts \cite{liu2011spin,liu2012spin,pai2012spin,zhu2018highly,hao2015giant,demasius2016enhanced,behera2022energy} to increase $\theta_{SH}$, but it came at the cost of increased resistivity, which increases the power consumption of the HM. We define the spin Hall power parameter($\epsilon_{SHE}$) that directly relates to the power consumption of the free FM switching process, accounting for the $\theta_{SH}$, $\rho$, and the thickness(t) of the HM. The $\epsilon_{SHE}$ defined in this work can be used to compare the power consumption of different SHE layers for the free FM switching/rotation via spin-orbit torque.


This paper is organized as follows. In Sec. \ref{design}, we give the details on the design of the ReLU circuit and introduce our spin Hall power factor. In Sec. \ref{simulation}, we describe our simulation platform where we couple Keldysh non-equilibrium Green's function formalism with Landau-Lifshitz-Gilbert-Slonzewski equations and the HSPICE circuit simulator to account for diverse physics of spin-transport and the CMOS elements in our proposed ReLU design. In Sec. \ref{results}, we present the results of our ReLU design and the performance of the proposed circuit against the thermal stability factor. Here we also show the result of our investigation into various heavy metals. 
We show that our design is resistant to thermal noise and that there exists a non-trivial power-error trade-off that leads to the energy-efficient circuit design using unstable free-ferromagnets.
In Sec. \ref{sec:cnn} we explore the potential application of our ReLU design using convolutional neural networks and show that our network achieves practically the same classification accuracy as ideal ReLU implementation.
We conclude in Sec. \ref{conclusion}.

\section{Design}
\label{design}

\begin{figure}[!t]
\centering
\subfloat[\label{fig:sche_3d}]{%
       \includegraphics[width=0.4\linewidth]{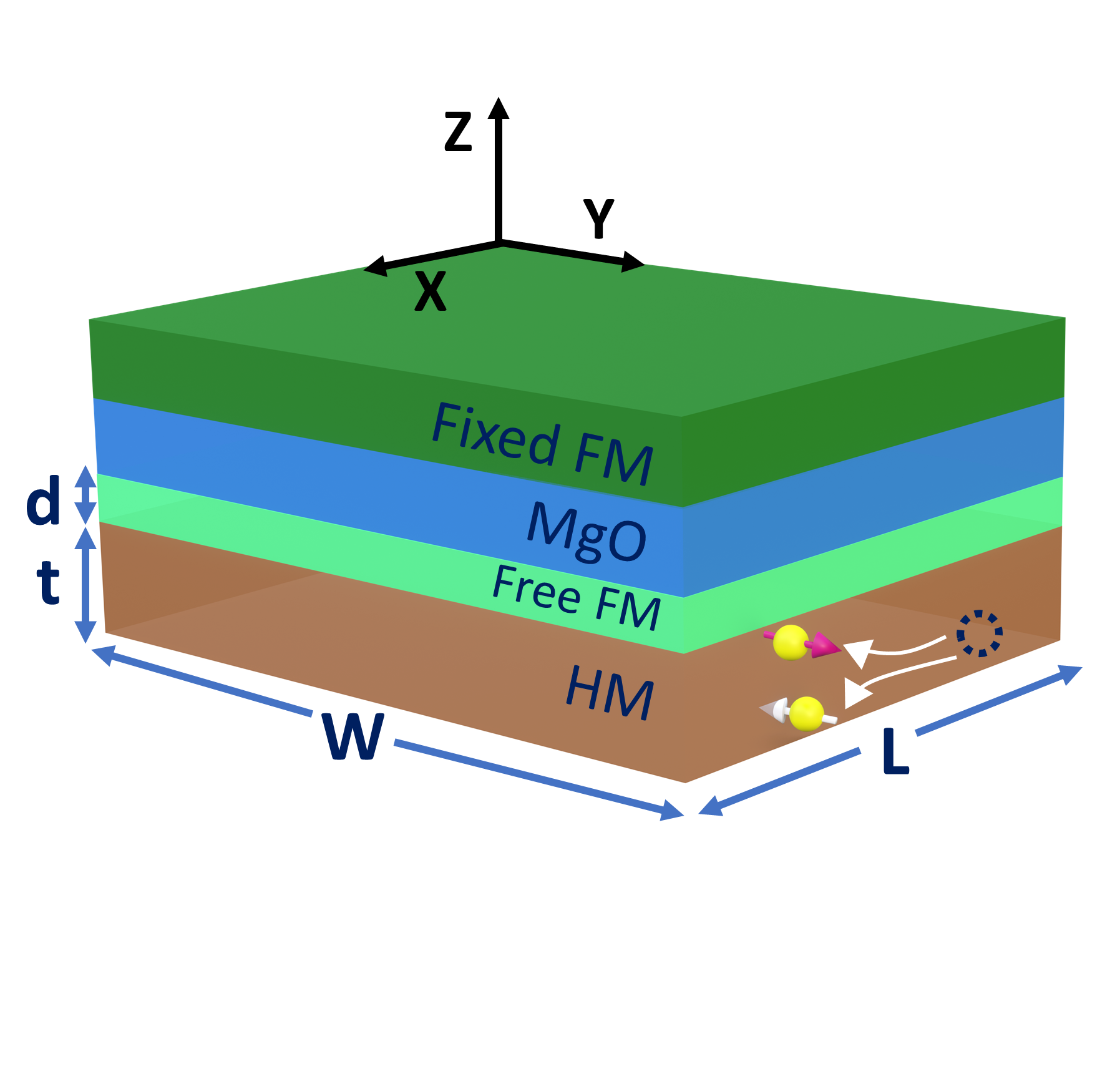}}
       \hfill
  \subfloat[\label{fig:relu_circuit}]{%
        \includegraphics[width=0.6\linewidth]{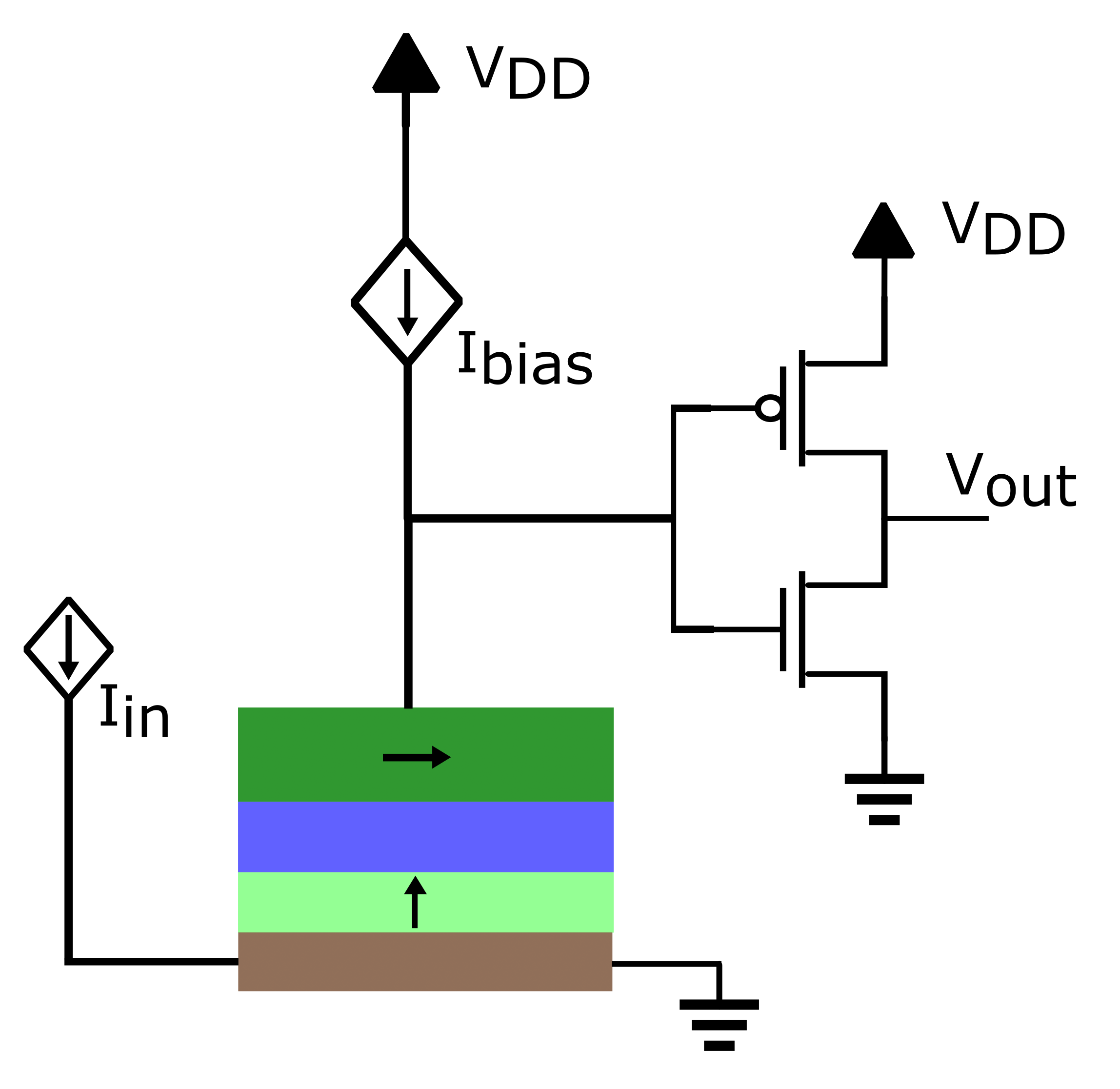}}
\caption{Design schematics. (a) The MTJ device is stacked on top of the heavy metal layer. Charge current is injected into the HM layer along $\hat{x}$, which injects $\hat{y}$-polarized spin current into the free-FM of the MTJ. (b). Circuit design for ReLU functionality. The current source $I_{bias}$ converts the change in resistance to a change in voltage that is connected to the CMOS inverter.}
\label{fig:circuit}
\end{figure}

\subsection{ReLU circuit}
MTJs are traditionally used as binary memories, here we show that an MTJ can be designed to have linear functionality. This is achieved when a spin current whose polarization is orthogonal to the anisotropy direction is applied to the free-FM. In this work, we inject a $\hat{y}-$ polarized spin current to the perpendicular magnetic anisotropy (PMA)-FM(CoFeB) in order to produce a linear rotation in $\hat{x}$- component of the magnetization as shown in Fig. \ref{fig:sche_3d}.

The linear rotation in magnetization is translated to the change in resistance via the TMR effect of the MTJ. The injected current($I_{bias}$) renders the resistance change to the voltage change across the MTJ, which drives the CMOS inverter to obtain the ReLU functionality, as shown in Fig. \ref{fig:relu_circuit}. The CMOS inverter operates in the linear region to invert and amplifies the voltage change across the MTJ, which results in ReLU function output as shown in Fig. \ref{fig:vout}. The current source $I_{bias}$ can be replaced with a resistor to obtain the ReLU functionality at the cost of decreased output ($V_{out}$) swing.

\subsection{Spin Hall power factor}
The SHE-driven MTJs are being explored extensively 
for low-energy applications using spintronic devices. Charge-to-spin conversion via spin Hall effect exhibited by the heavy metal is utilized for switching the free-FM layer in SHE-driven MTJs\cite{miron2011perpendicular,liu2012spin}. The power consumption of the switching can be minimized using a large charge-to-spin conversion factor (Eq. \ref{eq:Ic_Is}). There have been consistent efforts \cite{liu2011spin,liu2012spin,pai2012spin,zhu2018highly,hao2015giant,demasius2016enhanced,behera2022energy} to increase the spin Hall angle ($\theta$) through heavy metal engineering for enhanced charge-to-spin conversion factor. But an increase in $\theta $ usually comes with an increase in the resistivity of the heavy metal resulting in large power consumption.
Traditionally, the spin Hall conductivity\cite{behera2022energy} $\sigma_{SH}= \theta/\rho$ has been used to characterize the SHE, although this includes the effect of resistivity and spin Hall angle, it lacks analytical reasoning to represent power consumption of the SHE layer.
The spin Hall angle and resistivity also depend on the thickness of heavy metal \cite{hao2015giant, zhu2018highly}, compelling us to define a parameter that can unequivocally benchmark the various heavy metals for SHE switching power consumption.

The charge to spin conversion\cite{song2021spin,liu2012current,sengupta2017magnetic} of the SHE layer and the polarization and direction\cite{song2021spin,liu2012current,takahashi2008spin,hirsch1999spin} of the generated spin current are given by
\begin{equation}
    \theta _{SH} = \frac{J_s}{J_c}
\end{equation}

\begin{equation}
    I_s = \theta_{SH} \frac{L}{t} I_c 
    \label{eq:Ic_Is}
\end{equation}

\begin{equation}
    \hat{I_s} = \hat{I_c} \times \sigma
    \label{eq:she_dir}
\end{equation}
Here, $I_s$ is the spin current generated, $\theta_{SH}$ is the spin Hall angle of the heavy metal, L is the length of the heavy metal, t is the thickness of the heavy metal, and $I_c$ is the charge current injected. $\hat{I_s}$ is the direction of generated spin current flow, $\hat{I_c}$ is the direction of input charge current, and $\sigma$ is the polarization of the generated spin current. From Eq. \ref{eq:she_dir}, injection of charge current to heavy metal in $\hat{x}-$ direction results in y-polarized spin current injection to the free-FM (z-direction) on top of the HM layer.

The resistance (R) of the heavy metal is given by
\begin{equation}
    R=\rho \frac{L}{W t}
\end{equation}
Here, $\rho$ and W are the resistivity and width of the heavy metal respectively.

The power consumed by the heavy metal is given by $P_{HM}=I_c^2 R$. Here, $I_c$ can be written as
\begin{equation}
    I_c = \frac{I_s}{\sqrt{R V/d}} \frac{\sqrt{\rho t}}{\theta_{SH}} 
\end{equation}
where V, d are the volume and thickness of the free-FM layer. The power consumed by the heavy metal is given by
\begin{equation}
    P_{HM} = \frac{I_s ^2}{V/d} \frac{\rho t}{\theta_{SH} ^2} 
\end{equation}

Here the $I_s$ represents the spin current needed for switching the ferromagnet. For a given free-FM layer $I_s$, V and d are constants, so from the above derivation we define a spin Hall power factor  $\epsilon_{SHE}(= \frac{\sqrt{\rho t}}{\theta_{SH}})$ that can be used to compare different heavy metal. The material with the lowest $\epsilon_{SHE}$ will consume less power.

In our proposed circuit, $440 \mu A$ is required to achieve an output voltage of $0.35V$ for free-FM with a thermal stability factor of 45. The proposed factor is not limited to this work and can also be used in SHE-driven FM switching mechanisms. Some HMs such as Pt \cite{liu2011spin} affect the damping factor ($\alpha$) of the free layer as well, leading to an increase in the spin current ($I_s$) required. In such cases, the $\epsilon_{SHE}$ needs to be multiplied by the change in the damping factor ($\frac{\alpha_{old}}{\alpha_{new}}$). In our proposed ReLU circuit, the increased $\alpha$ has a negligible effect on the spin current requirement for the linear rotation.

\begin{table}
	\centering
\caption{ Simulation Parameters.}
	\label{table1}
	\begin{tabular}{|l| >{}m{3.5cm} | l |}
		\hline
\textbf{Symbol} & \textbf{Quantity} & \textbf{Value}\\ 
\hline
 $M_s$ & saturation magnetization & 1150 emu/$cm^3$ \\ 
 \hline
$H_k$ & anisotropy field & 1070 - 3670 Oe\cite{gajek2012spin} \\ 
 \hline  
 V & volume of ferromagnet & 1000 $nm^3$\\ 
 \hline
 d & thickness of the ferromagnet & 1 nm\\
 \hline
 $\Delta$ & thermal stability factor & 14.85 - 50.95 \\ 
 \hline
 $\alpha$ & Gilbert damping & 0.01\\ 
   \hline
$C_{MTJ}$ & MTJ capacitance & 26.56 aF \\
 
  \hline
$I_{bias}$ & bias current & 3.13 $\mu A$\\
\hline
$V_{DD}$ & voltage source & 0.5 V\\
\hline
$C_g$ & CMOS inverter input capacitance& 0.175 fF \\
\hline
$C_o$ & CMOS inverter output capacitance& 0.305 fF \\
  \hline
$\Delta  t$ & simulation time step & 0.5 pS\\
\hline
$\hbar$ & reduced Plank's constant & $1.055\times 10^{-34}$ Js\\
\hline
$k_B$ & Boltzmann constant & $ 1.38\times10^{-16} erg K^{-1}$\\
\hline
T & temperature & 300 K\\
\hline
	\end{tabular}
		
\end{table}

\section{Simulation Methods}
\label{simulation}

\begin{figure}[!t]
\centering
\subfloat[\label{fig:flow_a}]{%
       \includegraphics[width=0.6\linewidth,height=4.2cm]{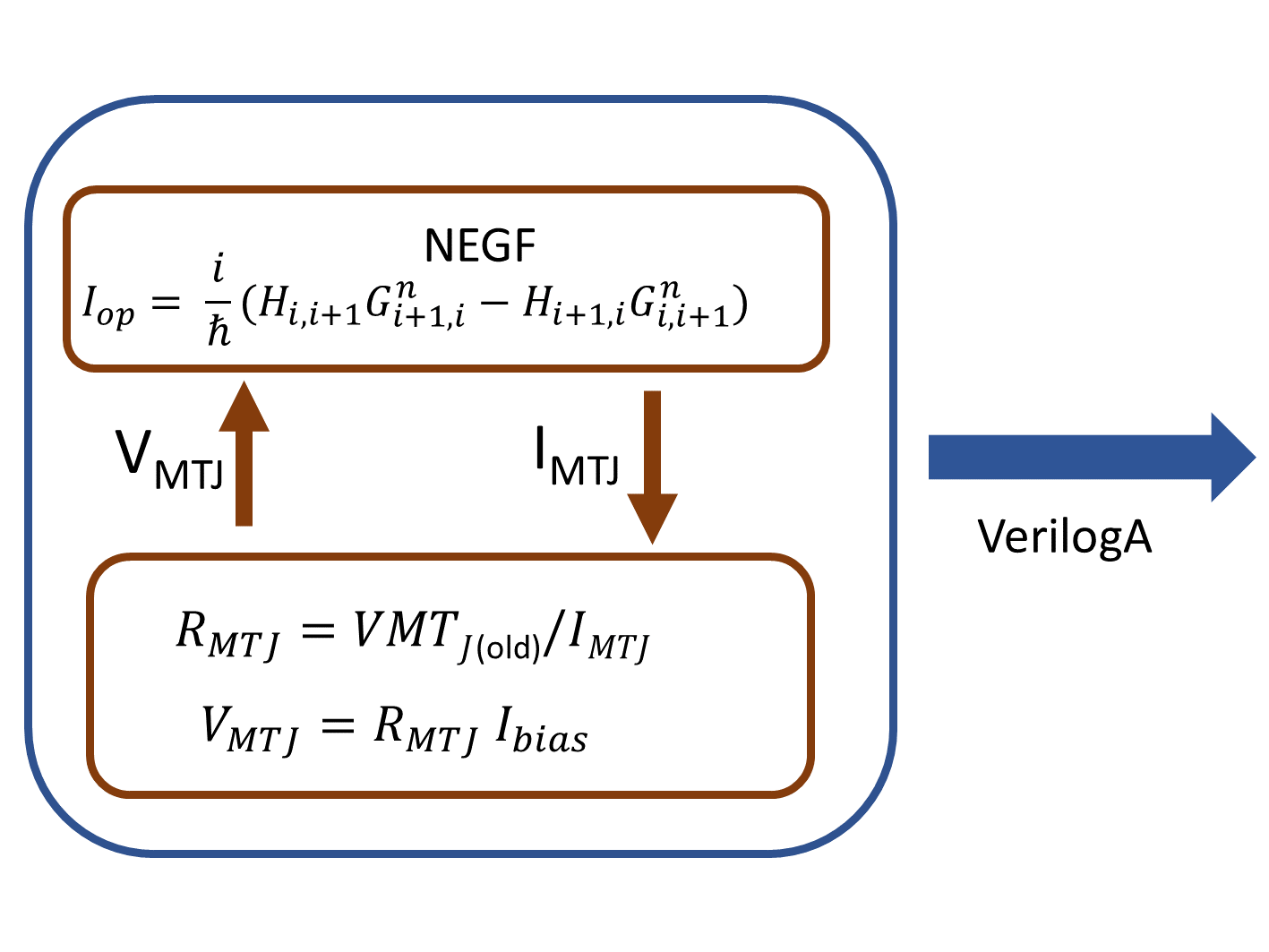}}
  \subfloat[\label{fig:flow_b}]{%
        \includegraphics[width=0.4\linewidth,height=4.2cm]{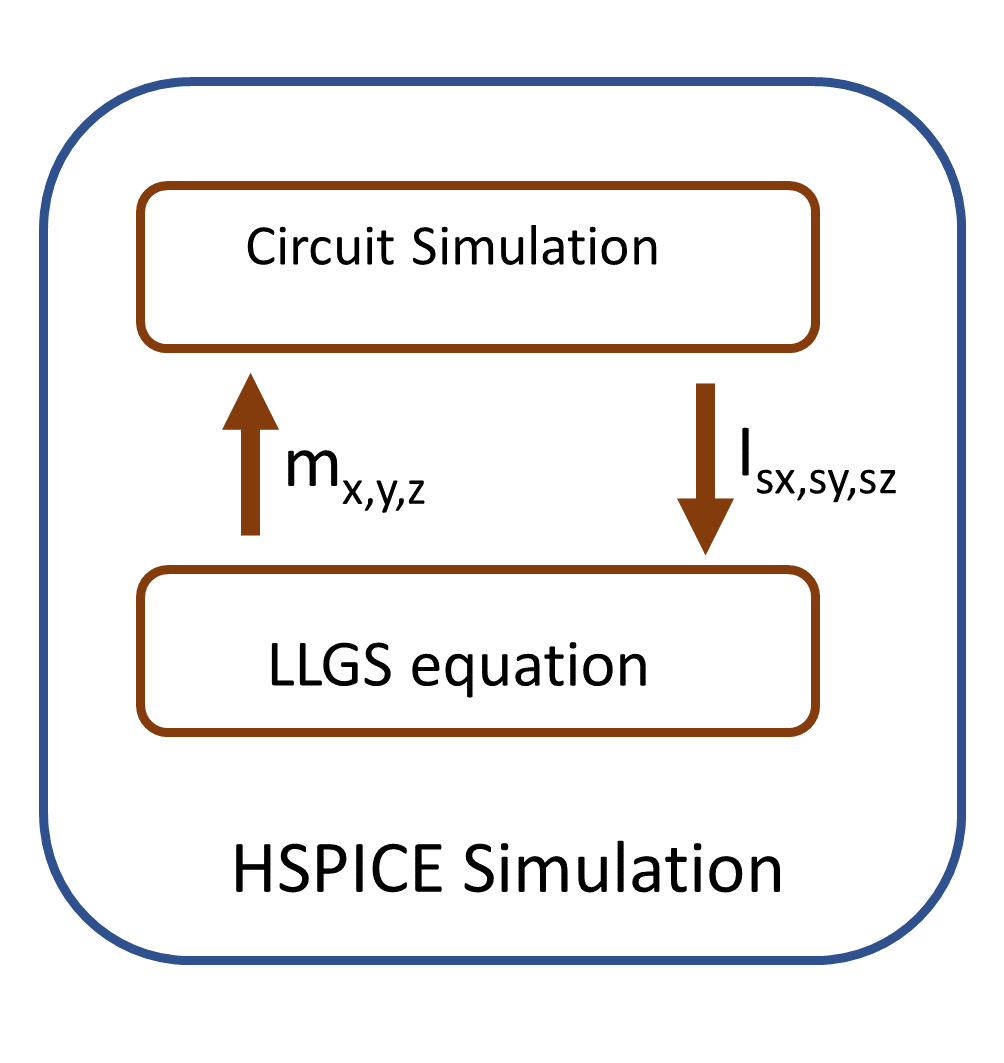}}
\caption{Hybrid NEGF-CMOS simulation platform setup. (a) The NEGF is self-consistently coupled with $I_{bias}$ to calculate the resistance of the MTJ. (b) The MTJ resistance is coupled to the HSPICE circuit simulator using VerilogA. The LLGS equation is coupled with other parts of the circuit to calculate free-FM magnetization.}
\label{fig:flow_chart}
\end{figure}

Figure \ref{fig:flow_chart} shows the schematic overview of the hybrid spintronics-CMOS simulation framework. The MTJ  and current source ($I_{bias}$) parameters are given to the NEGF simulator as shown in Fig. \ref{fig:flow_chart}. The NEGF simulator is self-consistently coupled with $I_{bias}$, since the device resistance depends on the MTJ angle and the voltage across the MTJ. The NEGF produces the device resistance vs MTJ angle plot. This result is coupled to the HSPICE circuit simulator via VerilogA. The circuit simulator simulates the entire circuit including the magnetization dynamics\cite{panagopoulos2013physics,sun2000spin} and it also simulates the approximation of the CMOS inverter pair based on a 16nm predictive technology model (PTM)\cite{Predicti66:online}.

\subsection{Quantum Transport}
We use the Keldysh NEGF technique \cite{camsari2020non,sharma2017resonant,datta2012modeling,datta2011voltage} to describe the transport through MTJ that has MgO sandwiched between free and fixed CoFeB FM layers. The NEGF formalism is given by

\begin{gather}
    G(E) = [EI-H-\Sigma]^{-1},\\
A(E) = i[G-G^\dagger],\\
\Gamma_{T,B} (E) = i([\Sigma_{T,B} (E)]-[\Sigma_{T,B} (E)]^\dagger),\\
\Sigma^{in}(E) = [\Gamma_T(E)]f_T(E) + [\Gamma_B(E)]f_B(E),\\
G^n= \int dE [G(E)] [\Sigma^{in}(E)][G(E)]^\dagger,\\
\Sigma=\Sigma_T + \Sigma_B
\end{gather}

Here $[H]$ is the device Hamiltonian, $[H]=[H_0]+[U]$, comprising device tight-binding matrix $[H_0]$ and the Coulomb charging matrix $[U]$, and $[I]$ is the identity matrix, $E$ is the energy variable. The charging matrix $[U]$ is calculated self-consistently using Poisson's equation. $G(E)$ is the Green's function matrix, $\Gamma_{T,B}, f_{T,B}, \Sigma_{T,B}$ are the broadening matrix, the Fermi function, and the self-energy matrices for the top (fixed) and bottom (free) FM layers respectively. $A$ is the spectral function, $\Sigma^{in}$ is the in-scattering function, and $G^n$ is the correlation matrix.

The quantum transport segment culminates with the calculation of the current operator ($I_{op}$) that represents the charge current between two lattice points i and i+1 is given by

\begin{equation}
    I_{op} = \frac{i}{\hbar}(H_{i,i+1}G^n_{i+1,i} - H_{i+1,i}G^n_{i,i+1})
\end{equation}

The current operator $I_{op}$ is $2 \times 2$ matrix in the spin space of the lattice point. Using this the charge current can be evaluated as

\begin{equation}
    I= q \int Real [Trace(\hat{I}_{op})] dE,
\end{equation}

where q is the quantum of electronic charge.

\begin{figure}[!t]
\centering
\subfloat[\label{fig:mag_res}]{%
       \includegraphics[width=0.5\linewidth]{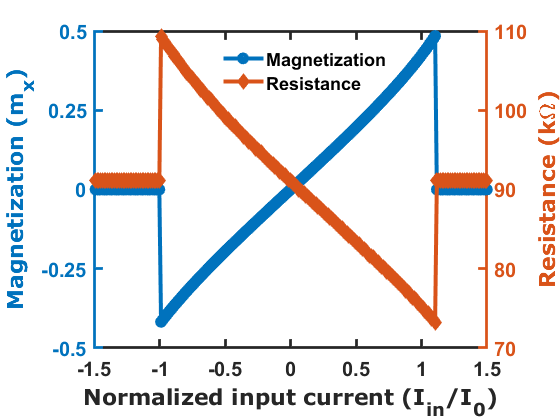}}
       \hfill
  \subfloat[\label{fig:vout}]{%
        \includegraphics[width=0.5\linewidth]{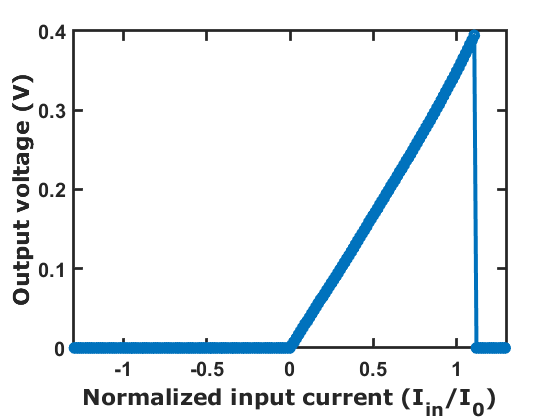}}
\caption{(a) The free-FM magnetization($m_x$) and the MTJ resistance as the normalized input ($I_{in}/I_0$) is varied with an initial magnetization oriented in the +z direction. $I_{in}$ is applied along $\hat{x}$. (b) The output of the ReLU circuit with varied normalized input. $I_0=380 \mu A$ and $\Delta = 45$ with $Au_{0.25}Pt_{0.75}$ heavy metal.}
\label{fig:relu}
\end{figure}

\subsection{Current injected MTJ}
In the ReLU circuit, we have employed the current source $I_{bias}$ to translate the change in MTJ resistance to voltage variation. The MTJ resistance depends not just on the free-FM magnetization but also on the voltage across MTJ \cite{vadde2022orthogonal}. The MTJ voltage itself depends on the resistance and the current source $I_{bias}$, so the device resistance and voltage need to be found self consistently. Figure \ref{fig:flow_chart}(a) shows the algorithm for this self-consistent calculation of the resistance while accounting for its voltage and current source $I_{bias}$ dependence. The self-consistent loop is run until the MTJ current is equal to the applied $I_{bias}$.

\subsection{Magnetization dynamics}
The LLGS equation \cite{slonczewski1996current,brataas2012current} is used to describe the magnetization dynamics of the free-FM. The LLGS equation is given by
\begin{multline}
  (\frac{1+\alpha^2}{\gamma H_k})\frac{d \hat{m}}{dt} = -\hat{m} \times \Vec{h}_{eff} - \alpha \hat{m} \times \hat{m} \times \Vec{h}_{eff} \\ - \hat{m} \times \hat{m} \times \Vec{i}_s + \alpha \hat{m} \times \Vec{i}_s,
\end{multline}
where $\hat{m}$ is the unit vector along the direction of magnetization of the free magnet, $\gamma$ is the gyromagnetic ratio, $\alpha$ is the Gilbert damping parameter, $\Vec{h}_{eff} = \frac{\Vec{H}_{eff}}{H_k}$ is the reduced effective field and $\Vec{i}_s= \frac{\hbar\Vec{I}_{s}}{2qM_sVH_k}$ is the normalized spin current. The term $\vec{H}_{eff}$ includes the contribution of the anisotropy field ($H_k$) and the thermal noise ($H_{th}$). The thermal noise is given by $ \langle H_{th}^2 \rangle = \frac{2\alpha k_B T}{\gamma M_s V}$ and $\langle \rangle$ represents the ensemble average \cite{sun2004spin}.

\section{Results}
\label{results}

\begin{table*}[b]
	\centering
\caption{Parameters for different HMs.} 
	\label{table2}
	\begin{tabular}{|l| >{}m{1.5cm} | >{}m{1.3cm} | >{}m{1.8cm} | >{}m{1.5cm} | >{}m{1.5cm} |   >{}m{2.5cm} | >{}m{1.5cm} |}
		\hline
\textbf{Heavy Metal} & \textbf{Spin Hall angle $\theta$} & \textbf{Resistivity $\rho$ ($\mu  \Omega cm$)} & \textbf{HM thickness t (nm)} & \textbf{HM Length  L (nm)} & \textbf{HM Width W (nm)} & \textbf{Spin Hall power factor $ \epsilon_{SHE} = \frac{\sqrt{\rho t}}{\theta}$ ($\sqrt{\Omega} nm$)} & \textbf{$I_0$ ($\mu A$)}\\ 
\hline
 $Pt$ \cite{liu2011spin} & 0.07 & 20 & 6 & 38.73 & 25.82 & 494.87 & 980 \\ 
 \hline
$\beta -Ta$ \cite{liu2012spin} & 0.12 & 190 & 4 & 10.26 & 97.46 & 726.48 & 1442 \\ 
 \hline  

 $\alpha + \beta -W$ \cite{pai2012spin} & 0.18 & 80 & 6.2 & 19.68 & 50.80 & 391.26 & 775 \\
 \hline
  $Au_{0.25}Pt_{0.75}$ \cite{zhu2018highly} & 0.3 & 83 & 4 & 15.52 & 64.42 & 192.06 & 380 \\
 \hline
  $\beta - W$ \cite{hao2015giant} & 0.35 & 210 & 9 & 14.64 & 68.31 & 392.79 & 778 \\ 
 \hline
  $W(O)$ \cite{demasius2016enhanced} & 0.5 & 200 & 6 & 12.25 & 81.65 & 219.09 & 434 \\ 
 \hline

$W_{0.88}Ta_{0.12}$ \cite{behera2022energy} & 0.58 & 260 & 5 & 9.81 & 101.98 & 196.58 & 389 \\
\hline

	\end{tabular}
\end{table*}

We show in Fig. \ref{fig:mag_res} the linear rotation in magnetization of the free FM layer of the MTJ. The linear rotation is achieved by injecting $\hat{y}-$polarized spin current into the free FM layer. The spin current is generated by applying charge current to the HM layer along $\hat{x}$. The TMR effect of the MTJ translates the linear magnetization change into a linear change in the resistance, this is shown in  Fig. \ref{fig:mag_res}. The resulting linear variation of MTJ resistance is employed in the circuit design (Fig. \ref{fig:circuit}) to realize the ReLU output as shown in Fig. \ref{fig:vout}. The output closely emulates the ReLU activation function for normalized inputs of less than 1. 
The parameters used in this design are given in Tab. \ref{table1}.

We evaluate the role of different HMs in our proposed ReLU design. The ReLU circuit's performance is assessed against the thermal stability factor ($\Delta=\frac{H_k M_s V}{2 k_B T}$) of the free FM layer. The $\Delta$ factor not only captures the stability of the free FM against thermal noise but also determines the spin current required for MTJ switching. We vary the $\Delta$ factor of the free FM by changing the anisotropy field while keeping the dimensions of the free FM fixed. A decrease in the $\Delta$ factor reduces the spin current needed for linear rotation, diminishing the HM's input charge current and power consumption.

\begin{figure}[]
\centering
    \subfloat[\label{fig:5she}]{\includegraphics[width=0.49\linewidth]{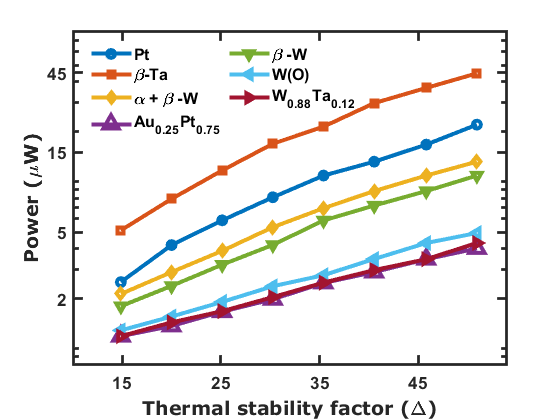}}    
    \subfloat[\label{fig:WO_power}]{\includegraphics[width=0.49\linewidth]{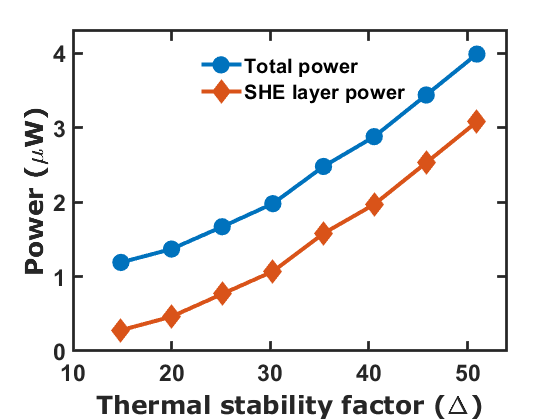}}
\caption{(a) Power consumption of the ReLU circuit with different HMs with varying $\Delta$. With y-axis in log scale (b) Power consumption of the entire ReLU circuit and power consumption of only the HM layer using $Au_{0.25}Pt_{0.75}$}
\label{fig_}
\end{figure}

The heavy metals used in this paper for the ReLU design are given in Tab. \ref{table2} along with respective parameters such as spin Hall angle $\theta$, resistivity $\rho$, and thickness $t$ taken from experimental works \cite{liu2011spin,liu2012spin,pai2012spin,zhu2018highly,hao2015giant,demasius2016enhanced,behera2022energy}. Using these parameters, the length $L$ and width $W$ are calculated such that the HM offers a resistance of 50 $\Omega$. The spin Hall power factor ($\epsilon_{SHE}$) and the normalizing current $I_0$ (current required to achieve maximum magnetization rotation with the ReLU output of 0.35V) are also shown in the table. It is inferred from Tab. \ref{table2} that the current $I_0$ accompanies  $\epsilon_{SHE}$. We can see from Fig. \ref{fig:5she} and Tab. \ref{table2}, that the average static power consumption of the ReLU circuit for different HMs also trails the $\epsilon_{SHE}$. The static power consumption decreases with $\Delta$ as it decimates the spin current required for the ReLU functionality. We show in Fig. \ref{fig:WO_power} the contribution of HM power consumption to that of the entire ReLU circuit for $Au_{25}Pt_{75}$ as the SHE layer. Along with the HM, MTJ consumes a fixed amount of power for translating the magnetization changes to voltage changes hence not affected by the change in $\Delta$. The MTJ sensing power consumption dominates the total ReLU power at lower $\Delta$s.

\begin{figure*}
    \centering
    \subfloat[\label{fig:power_err_pt}]{\includegraphics[width=0.25\linewidth]{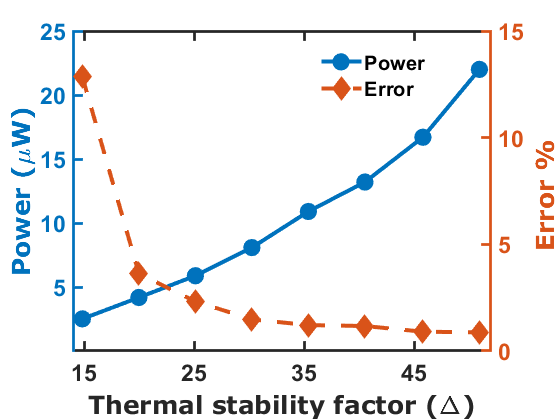}}
    \subfloat[\label{fig:power_err_beta_Ta}]{\includegraphics[width=0.25\linewidth]{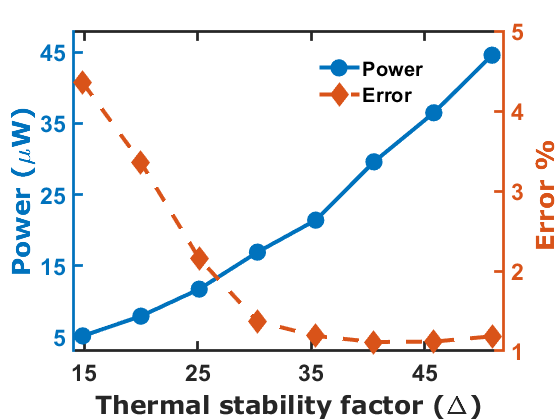}}
    \subfloat[\label{fig:power_err_alpha_beta_W}]{\includegraphics[width=0.25\linewidth]{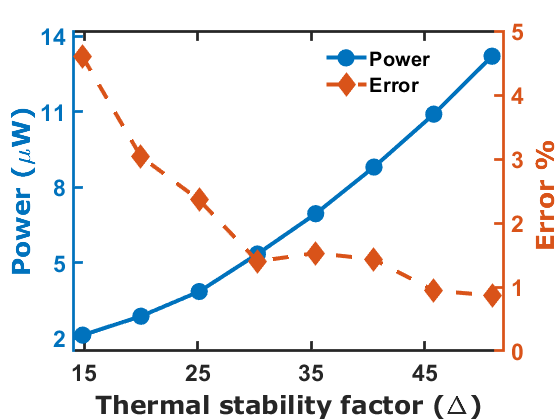}}
    \subfloat[\label{fig:power_err_AuPt}]{\includegraphics[width=0.25\linewidth]{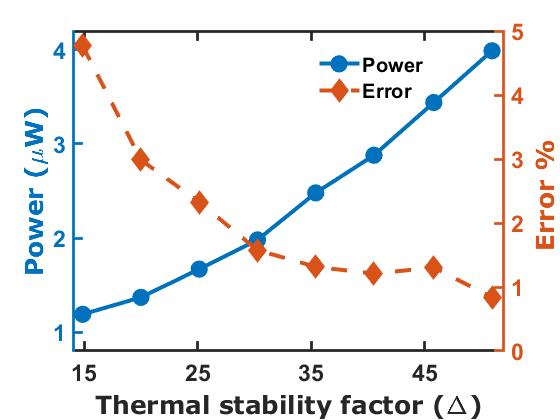}}\\
    \subfloat[\label{fig:power_err_beta_W}]{\includegraphics[width=0.25\linewidth]{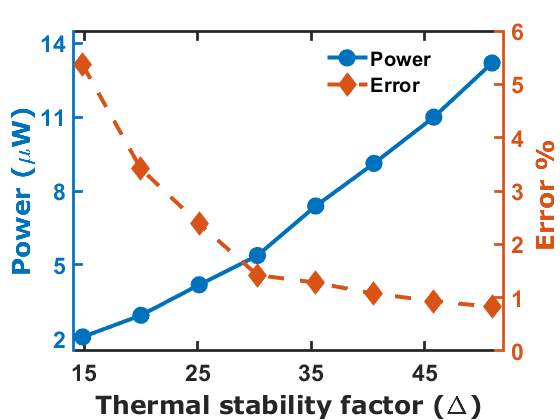}}
    \subfloat[\label{fig:power_err_oxy_w}]{\includegraphics[width=0.25\linewidth]{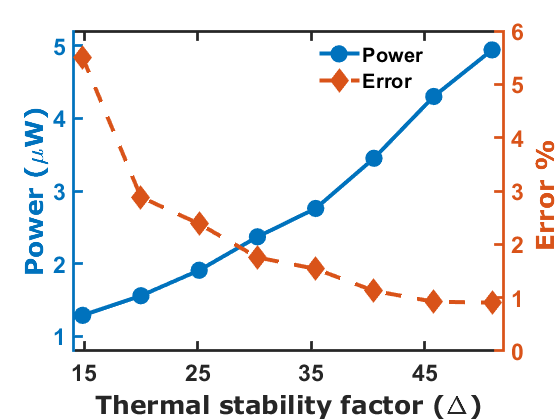}}
    \subfloat[\label{fig:power_err_WTa}]{\includegraphics[width=0.25\linewidth]{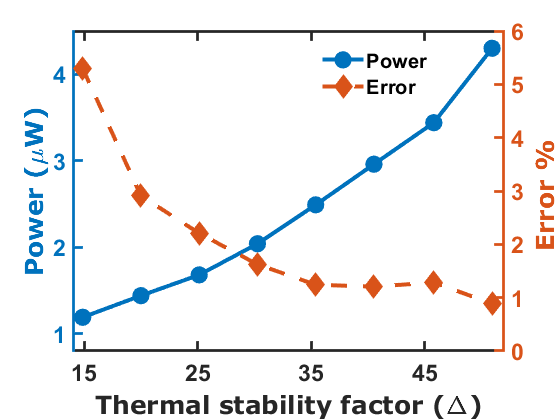}}
    
    \caption{Static power consumption of the ReLU circuit averaged over the entire input range, and the average absolute percentage error for a normalized input of 0.5 with (a) Pt, (b) $\beta - $Ta, (c) $\alpha + \beta -$W, (d) $Au_{0.25}Pt_{0.75}$, (e) $\beta - $W, (f) W(O), and (g) $W_{0.88}Ta_{0.12}$ as SHE layer.}
    \label{fig:power_err_all}

\end{figure*}

\begin{figure*}[b]
\centering
\includegraphics[width=0.98\linewidth]{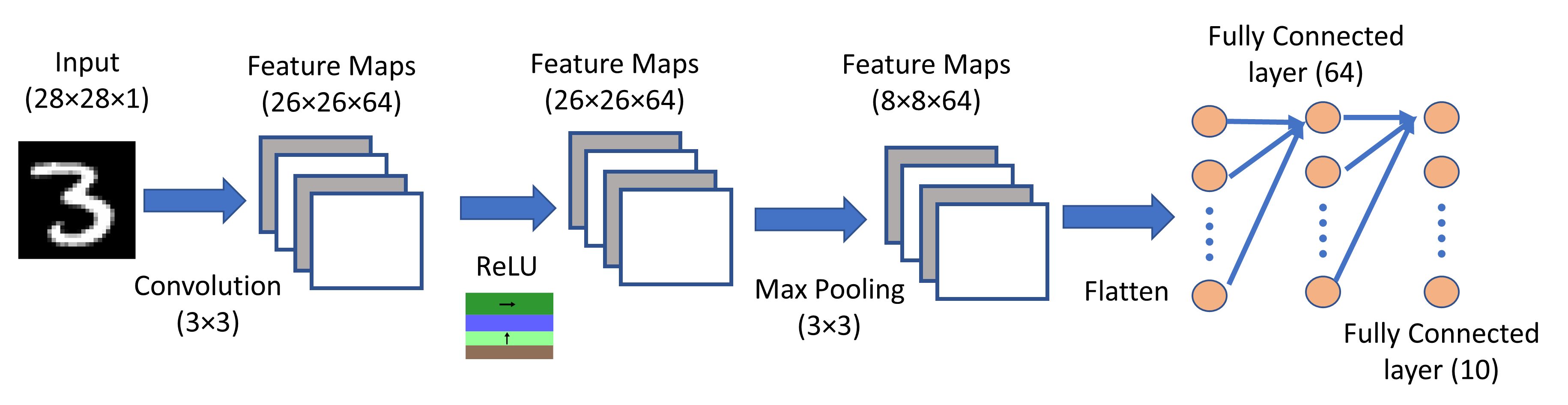}
\caption{Schematic of CNN architecture used for training in TensorFlow for the image classification task. The architecture has a $3\times3$ convolution layer followed by the ReLU activation function, max-pooling layers, and finally the fully connected layer. }
\label{fig:cnn_arch}
\end{figure*}

We show in Fig. \ref{fig:power_err_all} the static power consumption and the average absolute percentage error of the ReLU circuit for different HMs. The average of absolute error increases as $\Delta$ is decreased since the contribution of the thermal noise ($ \langle H_{th}^2 \rangle = \frac{2\alpha k_B T}{\gamma M_s V}$) increases in the effective magnetic field ($H_{eff}$). The $H_{eff}$ includes the effect from both thermal noise and the anisotropy field. As the $\Delta$ is reduced, the $ \langle H_{th}^2 \rangle$ stays constant, but the $H_k$ decreases. The effect of the thermal noise on the output is estimated using 100 Monte Carlo \cite{hammersley2013monte} simulations, with a normalized input ($I_{in}/I_0$) of 0.5. All the HMs show a decrease in power consumption and an increase in the error while the $\Delta$ is reduced, presenting an opportunity to optimize the circuit to consume less power while obtaining reliable results. For $Au_{0.25}Pt_{0.75}$ the static power consumption is 1.37 $\mu W$ while the absolute error percentage is $2.98 \%$ for $\Delta=20$ as shown in Fig. \ref{fig:power_err_AuPt}. These results suggest that an unstable ($\Delta<40$) free FM-based MTJ can be used to obtain reliable results for the ReLU circuit. 

\section{Application: Convolutional Neural Networks}
\label{sec:cnn}
Convolutional Neural Networks (CNN) are a class of artificial neural networks that produces excellent results in problems involving image data. Figure \ref{fig:cnn_arch} shows the architecture of the CNN used for classifying the MNIST and fashion MNIST data sets. Here we use our developed ReLU circuit instead of the ideal ReLU in the feature extraction stage of the CNN to train the network.

\begin{figure}[]
\centering
    \subfloat[\label{fig:digit}]{\includegraphics[width=0.49\linewidth]{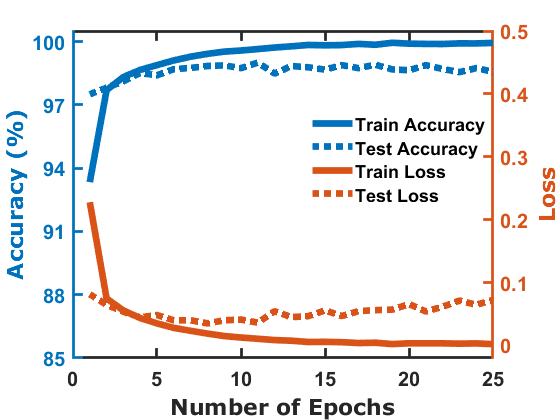}}    
    \subfloat[\label{fig:fashion}]{\includegraphics[width=0.49\linewidth]{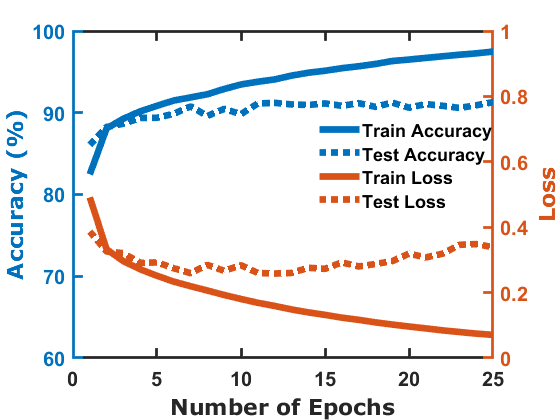}}
\caption{Classification accuracy(\%) and loss of the CNN as a function of epochs for different datasets: (a) MNIST (b) Fashion MNIST.}
\label{fig_}
\end{figure}

We show in Fig \ref{fig:digit} $\&$ \ref{fig:fashion} the accuracy and loss of the network during training and testing for MNIST and fashion MNIST datasets respectively. We notice an accuracy of $98.76\%$ on the test data for the MNIST dataset and an accuracy of $91.3\%$ for the fashion MNIST dataset. The accuracies for full software implementation(with ideal ReLU) are $98.86\%$ and $90.41\%$ for MNIST and fashion MNIST datasets. The accuracies using the non-ideal ReLU (our developed ReLU circuit) are closer to the ideal ReLU, this demonstrates the robustness of our ReLU circuit for neuromorphic applications. 

\begin{figure}[h]
\centering
\includegraphics[width=0.8\linewidth]{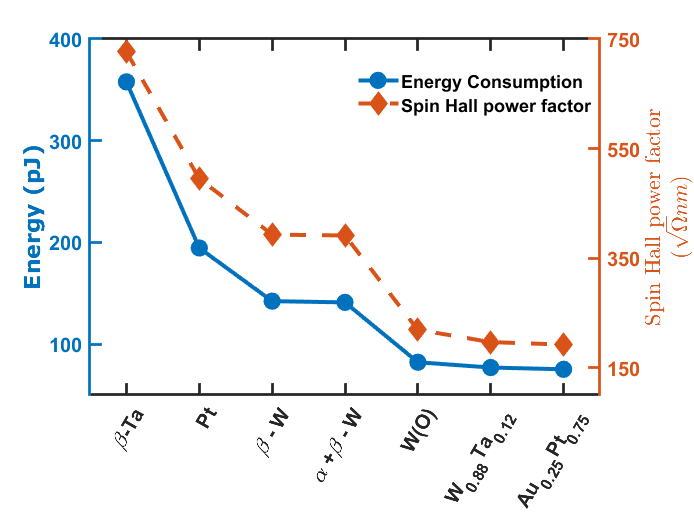}  
\caption{Energy consumption of the ReLU implementation in CNN architecture for testing a single sample for different heavy metals and the spin Hall power factors of the heavy metals.}
\label{fig:cnn_energy}
\end{figure}

Figure \ref{fig:cnn_energy} shows the energy consumption for ReLU implementation in our CNN architecture for different SHE layers, along with their spin Hall power factors. Here the free-FMs have a thermal stability factor of $20$. Here we see that the energy consumption of the ReLU implementation follows the spin Hall power factor. Our results suggest that $75 pJ$ of energy is consumed by the ReLU implementation in testing a single sample in our CNN architecture with $Au_{0.25}Pt_{0.75}$ as heavy metal and a free-FM with $\Delta=20$.

\section{Conclusion}
\label{conclusion}
In this paper, we showcased the linear rotation in the magnetization of free-FM and proposed a circuit design that effectively emulates the ReLU function, a fundamental component of deep learning neural networks. We introduced a new metric, the spin Hall power factor, to unequivocally quantify the SHE layers' power consumption. Our simulation results not only confirm the validity of this factor but also demonstrate its potential to significantly impact the design of SHE driven devices and circuits. We deploy our developed simulation framework that combines current injected MTJ with NEGF, LLG, and HSPICE circuit simulator, enabling us to design and analyze the proposed ReLU circuit with varying HMs. 

We demonstrated the existence of a non-trivial power error trade-off that enables the use of unstable free-FM for energy-efficient ReLU design. We demonstrated that the most energy-efficient realization of the proposed ReLU circuit consumes 1.37 $\mu W$ of static power with a low error rate of 2.98\% using the HM $Au_{0.25}Pt_{0.75}$ at $\Delta = 20$. Furthermore, we showed the potential of our ReLU design in CNNs, producing classification accuracies close to the software ReLU implementation with an energy consumption of 75 $pJ$ per sample.

\section*{Acknowledgements}
The author AS acknowledges the support of ISIRD phase-1 project of IIT Ropar. The author BM wishes to acknowledge the Science and Engineering Board (SERB), Government of India, for funding under the MATRICS grant (Grant No. MTR/2021/000388). 

\section*{Conflict of Interest}
The authors have no conflicts to disclose.


\section*{Data availability}
Data is available on request from the authors.

\bibliographystyle{IEEEtran}
\bibliography{references}

\end{document}